\documentclass[usenatbib]{mnras}
\usepackage{aas_macros}
\usepackage{natbib}
\usepackage[utf8]{inputenc}
\usepackage{graphicx,float}
\usepackage{amsmath}
\usepackage{amssymb}
\usepackage[dvipsnames]{xcolor}
\usepackage{soul}
\usepackage{hyperref}
\hypersetup{colorlinks=true,allcolors=teal}
\usepackage{microtype}
\usepackage[normalem]{ulem}
\usepackage{caption}
\usepackage{subcaption}
\usepackage{tabularx}
\usepackage{multirow}
\newcommand{\Mh}{\ensuremath{h^{-1}M_{\odot}}}
\newcommand{\Mpch}{\ensuremath{h^{-1}{\rm Mpc}}}
\newcommand{\kpch}{\ensuremath{h^{-1}{\rm kpc}}}

\newcommand{\avg}[1]{\ensuremath{\left\langle \,#1\, \right\rangle}}

\newcommand{\eqn}[1]{equation~\eqref{#1}}

\newcommand{\be}{\begin{equation}}
\newcommand{\ee}{\end{equation}}

\title[Redshift and cosmology dependence of mock halo catalogs]{Properties beyond mass for unresolved haloes across redshift and cosmology using correlations with local halo environment }
\author[Ramakrishnan \& Velmani]{
Sujatha Ramakrishnan$^{1}$\thanks{E-mail: rsujatha@iucaa.in}, 
Premvijay Velmani$^{1}$\thanks{E-mail: premv@iucaa.in}
\\  
 $^1$ Inter-University Centre for Astronomy \& Astrophysics,
      Ganeshkhind, Post Bag 4, Pune 411007, India}

\date{draft}

\begin{document}
\label{firstpage}
\pagerange{\pageref{firstpage}--\pageref{lastpage}}
\maketitle

\begin{abstract}
The structural and dynamic properties of the dark matter halos, though an important ingredient in understanding large-scale structure formation, require more conservative particle resolution than those required by halo mass alone in a simulation. This reduces the parameter space of the simulations, more severely for high-redshift and large-volume mocks which are required by the next-generation large sky surveys. Here, we incorporate redshift and cosmology dependence into an algorithm that assigns accurate halo properties such as concentration, spin, velocity, and spatial distribution to the sub-resolution haloes in a simulation. By focusing on getting the right correlations with halo mass and local tidal anisotropy $\alpha$ measured at $4 \times$ halo radius, our method will also recover the correlations of these small scale structural properties with the large-scale environment, i.e., the halo assembly bias at all scales greater than $5 \times$ halo radius.
We find that the distribution of halo properties is universal with redshift and cosmology. By applying the algorithm to a large volume simulation $(600\Mpch)^3$, we can access the $30-500$ particle haloes, thus gaining an order of magnitude in halo mass and two to three orders of magnitude in number density at $z=2-4$. This technique reduces the cost of mocks required for the estimation of covariance matrices, weak lensing studies, or any large-scale clustering analysis with less massive haloes.
\end{abstract}

\section{Introduction}
     The formation, growth, and evolution of galaxies and their host dark matter haloes are dictated by the hierarchical assembly of matter that exists in large-scale structures called the cosmic web \citep{1978MNRAS.183..341W}. This cosmic web environment has tidal influence over not only the properties of the dark matter halo \citep{2021MNRAS.508.1189C} but also has effects at the scale of individual galaxies. Many observations find that the color and specific star formation rate of galaxies are influenced by the proximity to dark matter filaments \citep{2018MNRAS.474..547K,2018MNRAS.474.5437L}. The intrinsic alignments between galaxies across large separations also due to the tidal field connecting them is an important systematic in the weak lensing cosmological analysis \citep{2002MNRAS.333..501B,2013MNRAS.432.2433H}. Thus there is good reason to incorporate the different measures of the tidal environment as well as their effects on tracers while building mock catalogs that simulate our universe. 
     
     From the observation side, the mock catalogs can assist survey optimisation, calibration, and other systematics and from the theoretical side, they provide test beds for exploring new physics \citep{2018ApJS..234...36M}.
     The amount of information contained in the different mocks is a trade-off between the volume and spatial resolution required, the time complexity which can be afforded, the level of approximations that can be tolerated in order to produce realistic mocks. The full-physics hydrodynamical simulations which are very accurate are computationally expensive to simulate large volumes \citep{2014MNRAS.445..175G,2015MNRAS.450.1937C,2019ComAC...6....2N}. On the other hand, we have semi-analytic galaxy formation models where galaxy properties are determined by a range of baryonic physics \citep{2015ARA&A..53...51S}, HODs and SHAMs which provide different prescriptions for the galaxy occupancy in the dark matter only simulations. More recently deep learning techniques have also been employed for the same \citep{2019arXiv190205965Z,2020MNRAS.495.4227K,2021PNAS..11822038L}.      As surveys become larger and demand unprecedented volumes of mock catalogs, the fast and practical prescriptions become an invaluable tool. 
     
     Some of the earliest and simplest mocks assume that the galaxy properties depend only on the mass of the host halo in the dark matter only simulation \citep{ss09,2011ApJ...736...59Z,2015MNRAS.453.4368G,2016MNRAS.457.4360Z,2018MNRAS.476..741D}. But there are several structural and dynamic properties distinguishing halos of the same mass; based on their assembly histories these haloes have different shapes, angular momentum, density profiles, etc. We already know, from simulations, the existence of `halo assembly bias' - the correlation between the large scale clustering or halo bias and the secondary properties of the halo beyond the halo mass \citep{2005MNRAS.363L..66G,wechsler+06,2008ApJ...687...12D,2010ApJ...708..469F}. Though this can consequently create `galaxy assembly bias' even in a model with a halo mass only assumption for galaxy occupancy, many studies with SAMs and hydrodynamical simulations confirm an additional signal that cannot be explained only with the association between galaxy properties and halo mass  \citep{2007MNRAS.374.1303C,2018ApJ...853...84Z,2021MNRAS.502.3242X,caz21}. 
Several models have incorporated effects of the halo properties such as halo concentration, accretion time in their galaxy mocks to tackle issues related to assembly bias and galaxy conformity \citep{hw13,masaki+13,pkhp15,2015MNRAS.452.4013K,2016MNRAS.460.2552H,2017ApJ...834...37L}.

The structural and dynamic properties of the halo also leave observational imprints in other ways. Many analytic models predict connections between halo angular momentum and galactic disc rotation \citep{1980MNRAS.193..189F,1998MNRAS.295..319M} and more recent attempts try to measure the spin of the halo from the baryonic components \citep{2021arXiv211011490O}. Possibilities of constraining other halo properties using observations are also explored \citep{2013MNRAS.432.1046B}, sometimes demonstrated using halo catalogs \citep{2021arXiv210105280B,2021arXiv210606656X}.

Since the particle resolution requirement to compute these halo properties is $10 \times$ larger than that required to compute halo mass, the accessible dynamic range gets reduced while incorporating such beyond halo mass effects.   
Large and poor resolution simulations run the risk of not only producing inaccurate small-scale properties, but the inaccuracies can also creep into the large-scale properties that are correlated with these small-scale ones. In \citet{2021MNRAS.503.2053R}, we demonstrated an algorithm that can be used to populate poorly resolved haloes with accurate mock halo properties that also bring out the correct assembly bias. The algorithm focuses on getting the right correlations between the internal properties and the anisotropy in the local tidal environment. This is motivated by many studies that point to the local cosmic web environment around a halo as being an important intermediary \citep{2017MNRAS.469..594B,2018MNRAS.476.4877M,2018MNRAS.476.3631P} which consequently drives the halo assembly bias \citep{2019MNRAS.489.2977R}. Here, we extend on the previous work to make the algorithm applicable to higher redshifts and other cosmologies.

The paper is organised as follows. Section~\ref{sec:methods} describes the simulations and halo properties used in this work. Section~\ref{sec:shuffling} motivates the rationale behind using the halo's tidal anisotropy $\alpha$ and mass $m$ \emph{at all available redshifts} as the primary properties to determine the distribution of other halo properties. This is done by showing that the assembly bias signal is unchanged even when halo properties are randomly shuffled around among halos of the same m and $\alpha$. Section~\ref{sec:fitting functions} provides, for five halo properties, the functional form of the distribution function whose parameters are fit with high-resolution simulation data from several $\Lambda \rm{CDM}$ cosmologies and redshifts ranging from 0 to 4.
In Section~\ref{sec:demo} we apply the algorithm to a several large-volume simulations and study the level of improvements in the mock catalogues that can be achieved at different redshifts and cosmologies. We summarise  in  Section~\ref{sec:summary}.  The Appendices provide some of the technical analyses relevant to the main text. The relevant codes produced for this work are available \href{https://github.com/rsujatha/mockhaloprop}{here}.

\section{Methods}
\label{sec:methods}
\subsection{Simulations}
\label{sec:simulation}
Our first suite of N-body simulations were performed using the \textsc{gadget-4} simulation code \citep{2021MNRAS.506.2871S} in different cosmological volumes but with the same number of particles ($1024^3$) in each of them; this includes one realisation of a large $600 \Mpch$ periodic box and three realisations of smaller $75 \Mpch$ boxes. We generate initial conditions for these simulations using $3^{\rm rd}$order Lagrangian perturbation theory at redshift $z=24$ using the \textsc{music-2} monofonIC code \citep{2020ascl.soft08024H,2021MNRAS.503..426H}; where we also use \textsc{camb} python library to compute the transfer function needed.
For this suite of simulations we consider $\Lambda$CDM cosmology with best fit parameters given by Planck's final data release \citep{2020A&A...641A...6P}.

In addition to these simulations, we also use another suite of N-body simulations previously used in \cite{pa20} and \cite{2021MNRAS.503.2053R}. They were performed with \textsc{gadget-2} code and initial conditions generated with \textsc{music} code using $2^{\rm{nd}}$ order perturbation theory. In particular, we use one simulation in a $200 \Mpch$ box with the same Planck 2018 cosmology, and two simulations of size $150 \Mpch$ and $300 \Mpch$ box in WMAP-7 cosmology \citep{2011ApJS..192...18K}.

\begin{figure}
    \centering
    \includegraphics[width=\linewidth]{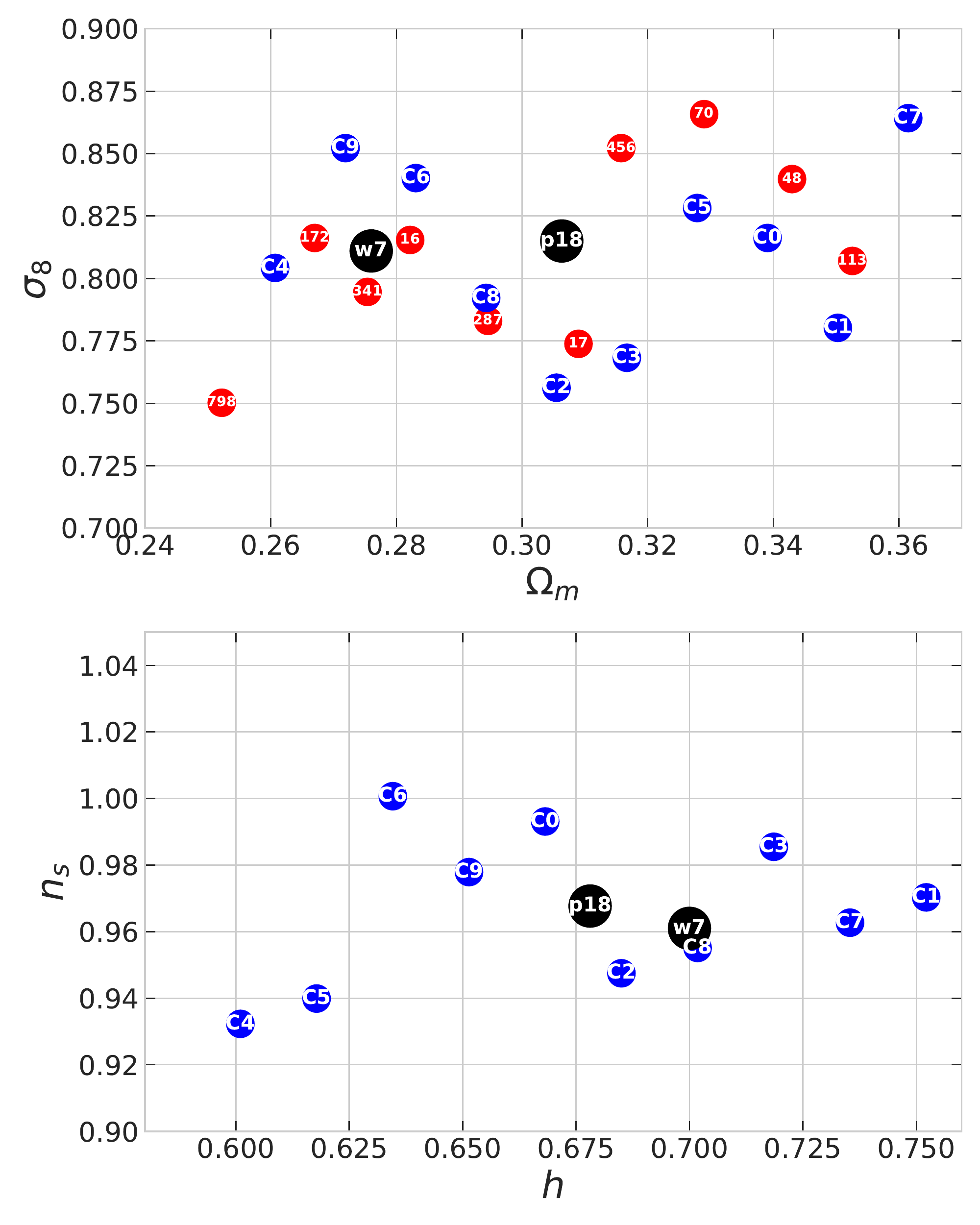}
    \caption{The set of cosmological parameters generated from latin hypercube for the simulations in SUITE-III is shown in blue, and the red text shows the cosmologies for the selected list of CAMELS-SAM simulations. All the CAMELS-SAM simulations have been run with $n_s=0.9624 ~\&~ h=0.6711$. For reference, the Planck 2018 (p18) and WMAP-7 (w7) cosmologies are also shown in black markers in the parameter space.}
    \label{fig:cosm-param-space}
\end{figure}

Our third suite contains nine large-volume simulations that were run on nine different non-standard cosmologies (see blue markers in the Figure \ref{fig:cosm-param-space}). This suite of simulations were also run with  GADGET-4 as in suite-I; each of these follows $512^3$ particles in $600 \Mpch$ periodic box.  The cosmological parameters were chosen by sampling the set $\{\Omega_m,\Omega_{b},h,\sigma_{8},n_{s}\}$ in a latin hypercube ranging 10$\sigma$ around the Planck 18 best fit values for all the parameters except $h$ for which we span $20 \sigma$ around the best fit value to accommodate the Hubble tension. We also use a set of 10 cosmologies from the CAMELS-SAMS suite \citep{2022arXiv220402408P} whose parameters also fall in the same range as discussed above for our third suite (see red markers in Figure~\ref{fig:cosm-param-space}).
A summary of the simulations used in this work is shown in Table~\ref{tab:sims}.

The haloes present in the simulation were identified and several structural properties were computed using the code \textsc{rockstar}\citep{2013ApJ...762..109B}, which uses Friends-of-Friends (FoF) algorithm in 6-dimensional phase space to identify haloes. Merger trees were constructed using the \textsc{consistent trees} code for two boxes ($200\Mpch$ and $150\Mpch$), where 201 snapshots equally spaced in the scale-factor between $z=12-0$ are available. To prevent contamination from spurious and non-virialised objects, only the haloes that satisfy the virial ratio cutoff $2T/|U|\leq 2$ are selected for further analysis (see \citet{2007MNRAS.376..215B}). The subhaloes are also discarded to control for the effects of substructure. All the simulations and analyses were run on the Pegasus cluster at IUCAA.



\begin{table*}
    \centering
    \begin{tabular}{c||c|c|c|c|c|c|c}
          &  $L_{\rm{Box}}$ $(\Mpch)$ & $N_{\rm{part}}$ & $m_{\rm{part}}$ $(10^9 \Mh)$ & $\epsilon_f$  $(\kpch)$ & $z_{\rm{init}}$ & $N_{\rm{real}}$ & Cosmology\\ \hline\hline & & & & & & &  \\
        \multirow{2}{*}{Suite-I} & 600 & $1024^3$ & 17.10 & 20 & 24 & 1  & \multirow{3}{*}{Planck 2018}\\ \cline{2-7}
         &  75 & $1024^3$ & 0.0334 & 2.44 & 24 & 3 & \\ \cline{1-7} 
        \multirow{3}{*}{Suite-II} & 200 & $1024^3$ &  0.63 & 6.5 & 99 & 1 & \\ \cline{2-8} 
         & 300 & $1024^3$ & 1.93 & 9.8 & 49 & 10 & \multirow{2}{*}{WMAP-7}\\ \cline{2-7}
         &  150 & $1024^3$ & 0.24 & 4.9 & 99 & 2 & \\
         \cline{1-8}
        Suite-III &  600 & $512^3$ &116-150 & 39& 24 & 1 & C1-C9$^{a}$\\
         \cline{1-8}
        CAMELS-SAM &  100 & $640^3$ & 0.26 - 0.37&1.3-2.6 & 127 & 1 & LH*$^{a}$
    \end{tabular}
    \\\hspace{\textwidth} {\footnotesize  $^{a}$The set of cosmologies considered are shown in Figure \ref{fig:cosm-param-space} }
    \caption{\textbf{List of simulations used:} The periodic box size of the simulation volume is denoted as $L_{\rm{Box}}$ and number of dark matter particles simulated is denoted as $N_{\rm{part}}$ while $m_{\rm{part}}$ denotes the mass of each of those particles. Here $\epsilon_f$ shows the comoving force softening length taken as one-thirtieth of the initial interparticle spacing \citep{2021MNRAS.504.3550G}. The number of realizations for each of those particular simulation configurations is denoted as $N_{\rm{real}}$ and $z_{\rm{init}}$ gives the initial redshift for those simulations.
    }
    \label{tab:sims}
\end{table*}


\subsection{Halo properties from a halo finder}
In this work, we will describe the halo's comoving radius by $R_{\rm 200b}$ which encloses an overall spherical overdensity of $200$ times the background matter density, and the associated mass is denoted by $M_{\rm 200b}$ or simply $m$. We will use the following structural and dynamic properties of the halo,
 \begin{itemize}
     \item Halo Concentration : $c_{\rm{200b}} = R_{\rm{200b}}/r_{s},$ where $r_s$ is the scale radius obtained by fitting for the NFW density profile \citep{1997ApJ...490..493N} .
     \item Halo Spin : $\lambda = J |E|^{1/2}/(GM_{\rm{vir}}^{5/2}),$ where $J$ is the halo’s angular momentum, $E$ is the sum of the halo’s potential and kinetic energies, and $M_{\rm vir}$ is its virial mass. The halo spin is dimensionless and quantifies the angular momentum of the dark matter halo \citep{1969ApJ...155..393P}.
     \item Halo shape ratio: $c/a$ quantifies ellipsoidal shape of the halo. It is obtained by first defining the mass ellipsoid $M_{i,j} = \sum_{n\in \rm{halo}} x_{i,n}x_{j,n}/r_{n}^2,$ where $x_{i,n},x_{j,n}$ are the components of the position of the $n^{\rm th}$ dark matter particle with respect to the center of mass of the halo and $r_{n}$ is the ellipsoidal distance to the center of mass. This tensor is computed iteratively, starting from a sphere of radius $R_{\rm vir}$ (the radius corresponding to $M_{\rm vir}$), each time including only particles inside an ellipsoid whose semi-major axis is $R_{\rm vir}$ and the semi-minor axes are defined by the eigenvalues of the tensor from previous iteration. The shape ratio $c/a$ is obtained after arranging the eigenvalues of the mass ellipsoid tensor as $a^2>b^2>c^2$. The iterative method allows the algorithm to adapt to the unknown shape of the halo \citep{2006MNRAS.367.1781A,2011ApJS..197...30Z}.
     \item Velocity ellipsoid ratio : $c_{v}/a_{v}$ is obtained by quantifying the triaxiality in the velocity dispersion of the halo by first constructing the velocity ellipsoid  $V_{i,j} = \sum_{n\in \rm{halo}} v_{i,n}v_{j,n}/N,$ where $v_{i,n},v_{j,n}$ are the $i^{\rm th}$ and $j^{\rm th}$ components of the peculiar velocity of the $n^{\rm th}$ dark matter particle with respect to the bulk peculiar velocity of the halo. We include only the particles inside the ellipsoid defined by the mass ellipsoid tensor calculated above. The ratio $c_{v}/a_{v}$ is obtained after arranging the eigenvalues of the velocity ellipsoid tensor as $a_{v}^2>b_{v}^2>c_{v}^2$.
     \item Velocity Anisotropy : $\beta = 1- {\sigma_{t}^2}/{2\sigma_{r}^2},$ where $\sigma_{t},\sigma_{r}$ are the tangential and radial dispersion of the halo \citep{1987gady.book.....B}. We compute $\beta$ only using particles inside the mass ellipsoid tensor computed above.
 \end{itemize}
The first three halo properties above were obtained as \textsc{rockstar} catalog output and the last two were calculated by modifying a local version of the same \citep{2019MNRAS.489.2977R}.
\label{subsec:haloprops}

\subsection{Large-scale environment}
In this work we focus on two large-scale properties.
\begin{itemize}
\item The two-point correlation function : We use the natural estimator of the correlation function \citep{1974ApJS...28...19P} which is given by,
\begin{equation}
    \xi (r) = DD(r)/RR(r)-1,
\end{equation}
where $DD$ is the number of halo pairs with separations in the range $(r,r+\Delta r)$ and $RR$ is same quantity for a random distribution of the same number density. For $N_{D}$ haloes in a periodic box, the random distribution $RR$ can be analytically expressed as $RR=N_{D}*(N_{D}/L_{\rm box}^3)4 \pi r^2 \Delta r$. This estimate of the correlation function is appropriate for our simulation boxes which are periodic.

\item The large scale bias: The estimator of the linear bias $b_1$ at large scales used here follows \citep{2018MNRAS.476.3631P} which computes a halo-centric bias estimate using appropriate weights, i.e.,
$\hat{b}_{1} = \sum_{{\rm low} k} N_{k} P_{\times}(k)/\sum_{{\rm low} k} N_{k} P(k)$. Here $P_{\times}(k)$ is the halo-matter cross power spectrum taking one halo at a time, $P(k)$ is the matter power spectrum, $N_{k}$ is number of $k$ modes in each $k$ bin. This is the least squares estimator under the assumption  of  Gaussian  errors  when  the  number  of haloes in the cross-power calculation is small \citep{pa20}. 

When averaged over a sufficiently large number of haloes, the halo-centric bias is equivalent to the traditional estimate, for e.g., the ratio of the halo-matter cross-power spectrum to the matter-matter power spectrum for small k modes ($\leq 0.1 \Mpch$)
\footnote{In the estimation
of bias by any method (traditional or halo-centric), the computationally costliest step is taking Fourier
transforms in order to obtain the power spectrum. The advantage of using halo-centric bias is that it requires us to compute this costly step only once after which we can obtain the bias of an arbitrary halo population by
simply taking the mean of previously computed halo-centric bias. In contrast, the traditional estimate would
require us to compute Fourier transforms every time we need to find the bias of an arbitrary halo population. The analyses here deal with the bias of several subsets of halo populations and hence can be obtained quickest with halo-centric bias.}.
\end{itemize}

\subsection{Local Environment : Tidal anisotropy }
The local environment around a halo is described by its tidal anisotropy parameter which characterises the anisotropy of the tidal forces experienced by the halo \citep{2018MNRAS.476.3631P}. It is expressed in terms of the eigenvalues $\lambda_1,\lambda_2,\lambda_3$ of the tidal tensor as follows:
\be
\alpha \equiv \sqrt{q^2}/(1+\delta),
\ee
where $q^2 = 1/2[(\lambda_{1}-\lambda_{2})^2+(\lambda_{2}-\lambda_{3})^2+(\lambda_{3}-\lambda_{1})^2]$ is the tidal shear and $\delta = \lambda_{1}+\lambda_{2}+\lambda_{3}$ is the matter overdensity.

The overdensity field is computed in a $1024^3$ grid using the cloud-in-cell (CIC) algorithm which is then smoothed with a Gaussian kernel of scale size $R_{G}$. We make copies of the smoothed density field corresponding to 50 logarithmically spaced smoothing scales $R_{G}$ spanning the different sizes of haloes in the simulation box; in Fourier space each smoothed field would be $\delta(\mathbf{k};R_{G}) = \delta(\mathbf{k}) e^{-\mathbf{k}^2R_{G}^2/2}$. The tidal tensor field corresponding to every $R_{G}$ can be obtained by inverting the Poisson equation and taking derivatives, i.e, the inverse Fourier transform of $(k_{i}k_{j}/k^2)\delta(\mathbf{k};R_{G})$. The tidal anisotropy $\alpha$ for each halo is obtained by evaluating the eigenvalues of the tidal tensor field at the nearest gridpoint to the center of the halo. The tidal tensor for computing $\alpha$ for each halo is obtained by interpolating between the two pre-computed scales $R_{G}$ that are closest to $4R_{\rm 200b}/\sqrt{5}$\footnote{This scale is the Gaussian equivalent of a Tophat scale of $4 \times R_{\rm 200b}$(see Appendix A2 of \citet{2018MNRAS.476.3631P}) .}. The tidal anisotropy $\alpha$ at this scale has the maximum correlation with $b_{1}$ and least correlation with $\delta $\citep{2018MNRAS.476.3631P}.
Later in section~\ref{sec:fitting functions} where the method for creating our mock halo catalog is described, we will use the standardised or normalised tidal anisotropy $\tilde{\alpha}$,
\be
\tilde{\alpha} = \dfrac{\ln \alpha - \avg{\ln \alpha|m}}{\sqrt{\rm {\rm Var} (\ln \alpha|m)}}.
\ee
Here, $\avg{\ln \alpha|m}$ and $\sqrt{\rm {\rm Var} (\ln \alpha|m)}$ are the mean and central $68.3$ percentile of $\ln \alpha$ in narrow bins of mass $m$ \footnote{ In the simulations, we compute $\avg{\ln \alpha|m}$ and $\sqrt{\rm {\rm Var} (\ln \alpha|m)}$ in mass bins and use this data to interpolate its value to any choice of mass as required in the computation of $\tilde{\alpha}$ for a halo.}.  In the standardised form, $\tilde{\alpha}$ has a normal distribution (see Figure 1 of \citealp{2020MNRAS.499.4418R}) which will later simplify our search for the conditional distribution of the halo property given the nature of the halo's local environment (i.e., $\alpha$).

\begin{figure*}
\includegraphics[width=1.0\textwidth]{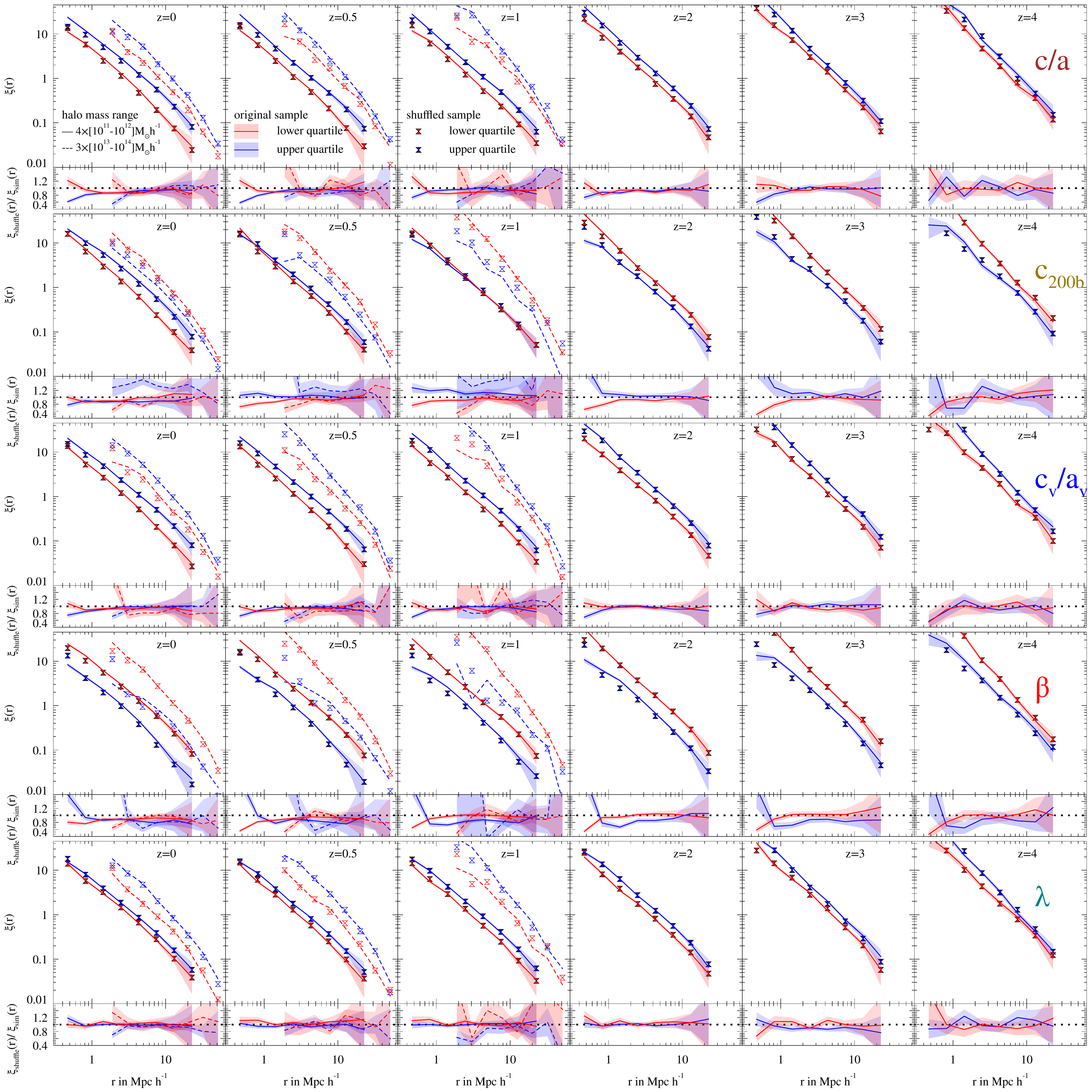}
\caption{\textbf{Real-space 2-point correlation function $\xi(r)$}. Comparison of $\xi(r)$ of the haloes measured in the simulations (lines) with that obtained from the same haloes shuffled in bins of tidal anisotropy (markers) as described in section~\ref{sec:shuffling} in two narrow mass bins. The solid (dashed) lines and the filled (empty) markers show this comparison corresponding to the mass ranges $4\times10^{11}-4\times10^{12}\,(3\times10^{13}-3\times10^{14})\,\Mh$. Each row shows the halo clustering in quartiles of a single halo property and different panels in each row corresponds to different redshifts as indicated by the labels. For halo shape $c/a$, spin $\lambda$ and velocity ellipsoid asphericity $c_v/a_v$, the upper quartile is more clustered than the lower quartile. For velocity anisotropy $\beta$, the lower quartile is more clustered than the upper quartile while the halo concentration $c_{\rm 200b}$ shows opposite trends depending on the chosen mass range and redshift. For each quartile of each property, the shuffled sample accurately reproduces  $\xi(r)$ at separations $5R_{\rm 200b}^{min}\lesssim r/\Mpch\lesssim45$ where $R_{\rm 200b}^{min}$ is the radius of the smallest halo in the sample. This can be seen quantitatively in the lower panels where the ratio of $\xi(r)$ in the shuffled samples to the original sample is consistent with $1$. We have used 64 jackknife samples to obtain errors for both the higher and lower mass samples which were obtained from the $600\Mpch$ and $200\Mpch$ boxes from Suite-I respectively. }
\label{fig:2ptcorr}
\end{figure*}

\begin{figure*}
 \includegraphics[width=0.95\linewidth]{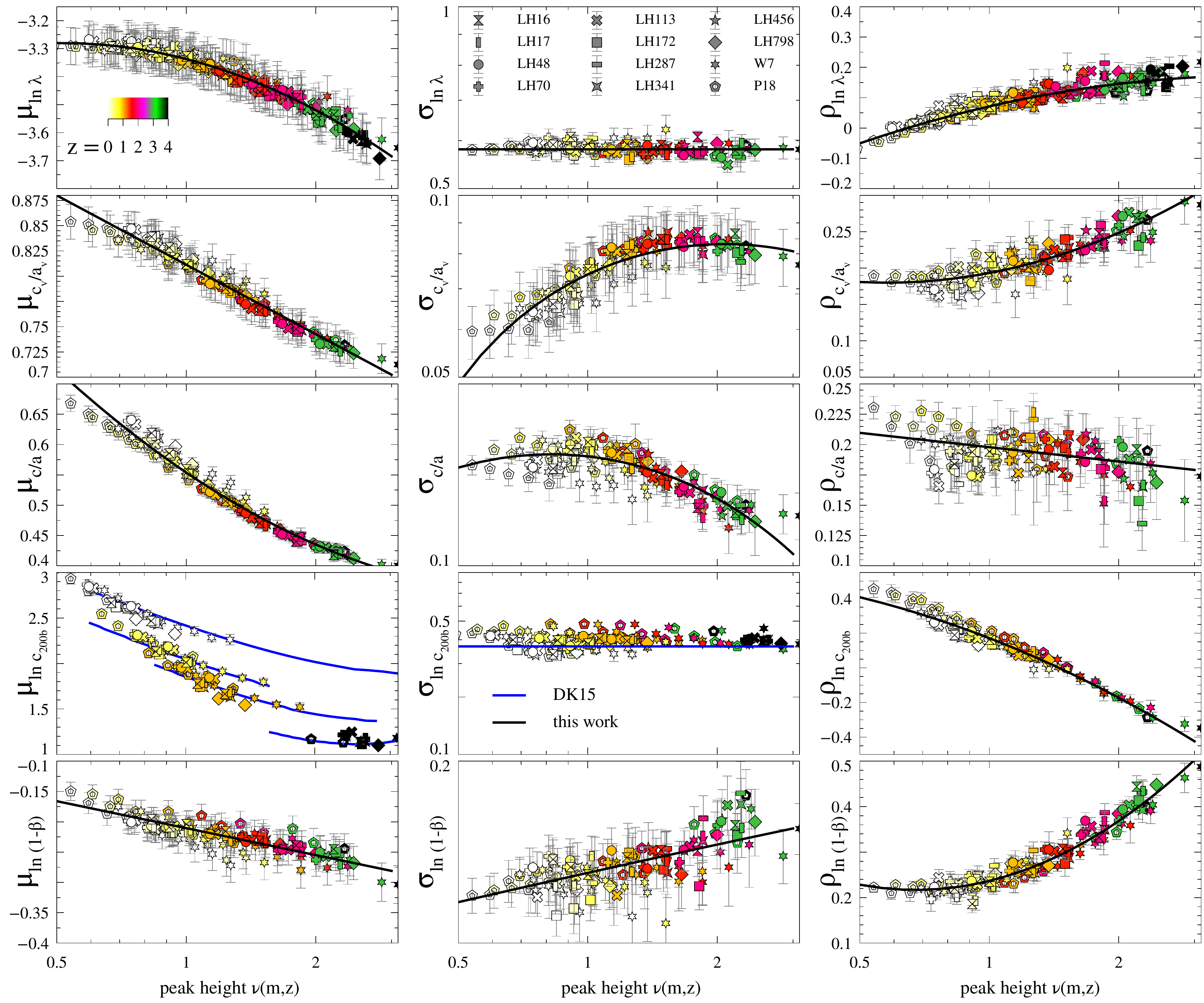}
 \caption{\textbf{Summary statistics of the halo properties as a function of peak height:} median halo property $\mu_{c}$ \emph{(left panel)},  central 68.3 percentile of the halo property $\sigma_{c}$ \emph{(middle panel)} and the correlation coefficient $\rho_{c}$ \emph{(right panel)} of the halo property {$c$} with standardised tidal anisotropy $\tilde{\alpha}$ . Each row pertains to a halo property as per the y axis and different colours are used to denote data from redshifts ranging from $0$ to $4$ as per the colorbar. Different markers show different cosmologies probed with high-resolution boxes as indicated in legend and detailed in Table~\ref{tab:sims} and Figure~\ref{fig:cosm-param-space}. Solid lines show polynomial fits to the data in the variable $\ln\nu$ (Tables~\ref{table:ca}-\ref{table:c200b}), except for $\mu_{\ln c_{\rm 200b}}$ and $\sigma_{\ln c_{\rm 200b}}$ where we show the calibrations from \citet[][DK15]{2015ApJ...799..108D}. These curves contain all the information required to generate realistic halo properties endowed with accurate cosmology and redshift dependent halo assembly bias, by sampling the probability distribution function $p(c|m,\tilde{\alpha})$ from \eqn{eq:probdist} which can be constructed knowing $\mu_c$, $\sigma_c$ and $\rho_c$ for each property $c$.} 
 \label{fig:fits}
\end{figure*}
\section{Shuffling Exercise}

\label{sec:shuffling}
In this section, we perform a shuffling exercise that explains the premise of our mock-making algorithm. It closely follows section 3.2 of \citet{2021MNRAS.503.2053R} where they shuffle the halo properties in bins of mass and tidal anisotropy before computing the two-point correlation function. This exercise will randomize or erase any associations that may exist between the halo's spatial location and its internal properties except those that linked through mass and tidal anisotropy. The specifics of the procedure are given below. Though the assembly bias signal can be more concisely understood in terms of the peak height\footnote{The peak height is defined as $\nu(m,z)=\delta_{c}(z)/\sigma(m)$, where $\delta_{c}$ is the critical threshold for spherical collapse and $\sigma(m)$ is the standard deviation of linear fluctuations smoothed with a spherical Tophat kernel at the Lagrangian radius scale. Both the numerator and the denominator in the definition are linearly extrapolated to $z=0$.} (see section~\ref{sec:fitting functions}), here we show the explicit trends with mass and redshift. 

We choose the haloes in two mass ranges -  $4\times10^{11}-4\times10^{12}\,(3\times10^{13}-3\times10^{14})\,\Mh$ obtained from the $200 \Mpch$ box and $600 \Mpch$ box, respectively. The haloes within these mass bins are divided into five quintiles of the tidal anisotropy parameter $\alpha$. The halo properties are then randomly shuffled within each of these quintiles. We compute the two-point correlation function in the upper and lower quartiles of the halo property in the shuffled and original-unshuffled sample and compare them.\footnote{The numpy percentile routine can provide q-th percentile of any data. We use this to select haloes whose halo property $c$ lies in the upper quartile (greater than $75^{th}$ percentile of $c$) and lower quartile (lower than $25^{th}$ percentile of $c$). Alternatively, we  can also first rank order the haloes according the magnitude of their halo property, then choose the top $1/4^{th}$ of the population(upper quartile) and bottom $1/4^{th}$(lower quartile) according to this ordering.   } The above procedure is repeated for all the available redshifts i.e, $z = 0,0.2,0.5,1,1.5,2,3,4$ except for the higher mass range at redshifts $z>1$ where the number of haloes is too few to estimate the two-point correlation. Figure~\ref{fig:2ptcorr} shows the resulting match between the original simulation and the shuffled samples for six redshifts. The shuffled sample not only matches the original sample well at a single redshift but also captures the time evolution of the halo assembly bias up to redshift $z=4$. For e.g., there is a flip or inversion of the clustering strength of the higher concentration compared to the lower concentration haloes as we go from current to earlier redshifts and the shuffled sample also exhibits this flip which can be seen at $z=1$ for the lower mass range. Consistent results have been verified for other intermediate redshifts also but we omit showing them for brevity. 

One detail that is overlooked in the previous discussion is that the two-point correlation of the shuffled sample does not show the assembly bias split in the clustering strength and does not match with the simulations for very small separations, specifically, lesser than five times the radius of the smallest halo ($5\times R_{200b}^{min})$ which can be clearly seen in the ratio between $\xi(r)$ of the shuffled mocks and original simulations shown in the bottom panel of each main plot. 
At these separations, the tidal anisotropy calculation in a radius of $4R_{200b}$ of each of the halo pairs will include the physical presence of the other. This close presence of a neighbor may influence halo properties in more ways than merely a tidal effect  \citep[see, e.g.,][for the influence of neighbors on assembly bias]{2019MNRAS.486.1156J,2018MNRAS.475.4411S}. 

Overall, this section provides us sufficient motivation for describing the sampling distributions of halo properties to be conditioned only on halo mass and local environment and hints at why mocks made with such a sampling may likely be successful at reproducing the clustering signal at all redshift considered up to $z=4$.

\section{Fitting Functions for higher redshifts and other cosmologies}
\citet{2021MNRAS.503.2053R} showed that sampling probability distribution of the halo's internal property $c$ (which is Gaussian distributed) can be modelled with a conditional distribution having the following parameters,
\label{sec:fitting functions}
\begin{align}
p(c|\tilde{\alpha})&=\dfrac{e^{-(c-\sigma_{c}\rho_{c} \tilde{\alpha}-\mu_{c} )^2/2\sigma_{c}^2(1-\rho_{c}^2)}}{\sqrt{2 \pi\sigma_{c}^2(1-\rho_{c}^2)}}\,.
\label{eq:probdist}
\end{align}
Here, $\mu_{c}$ and $\sigma_{c}$ are the mean and standard deviation of the marginal distribution $p(c)$ and $\rho_{c}$ is the correlation coefficient between $c$ and $\tilde{\alpha}$. However, these parameters were calibrated only at redshift $z=0$ using the WMAP-7 simulation. In this section, we will refit these parameters of the probability distribution with measurements from simulation snapshots of varying redshifts ($z=0,0.2,0.5,1,1.5,2,3,4$) and 12 different cosmologies. As the trends are more universal with peak height than halo mass, we will continue to fit these parameters as a function of $\ln \nu$. This point is most apparent in the correlation coefficient $\rho_{c_{200b}}$ which changes from positive to negative with increasing mass or peak height (see Figure~\ref{fig:c200brho}). In the case of other parameters, it is not so apparent whether universal relation exists with peak height or mass as they are almost constants with less variability. Overall we can conclude from the right panel of Figure~\ref{fig:fits} that there is negligible dependence of assembly bias ($\rho_c$) on cosmology and redshift. The evolution of assembly bias has been previously studied in \citet{2007MNRAS.377L...5G,2019MNRAS.484.1133C} and parametrised in \citet{wechsler+06} and negligible dependence on cosmology has been reported \citep{2021MNRAS.507.3412C} consistent with the findings in this section. 

Figure~\ref{fig:fits} shows the median halo property $\mu_{c}$, central 68.3 percentile of the halo property $\sigma_{c}$\footnote{The $\sigma_{c}$ defined here may not be confused with standard deviation of density fluctuations defined at various scales i.e., $\sigma(m)$ and $\sigma_{8}$. } and the correlation coefficient $\rho_{c}$ of the halo property with standardised tidal anisotropy $\tilde{\alpha}$ in bins of peak height.  We use 256 jackknife samples within each simulation box to estimate errors in these measurements. There are systematic offsets in measurements even with well-resolved haloes between the larger box and smaller box of the same redshift and cosmology. To account for this as well as other binning and numerical errors, we add in quadrature $2\%$ of the binned measurement to its error \citep{2019ApJ...871..168D} with a few exceptions; for the parameters that are already numerically close to zero at some or all peak heights, i.e, $\rho_{\ln c200b},\rho_{\ln \lambda},\sigma_{c_{v}/a_{v}},\mu_{\ln (1-\beta)}$, we add to the error $10\%$ of the measurement in the peak height bin. Tables~\ref{table:ca}-\ref{table:c200b} provide the fitting parameters as a function of $\ln \nu$ and the degree of the fit is chosen after an analysis of the Akaike Information Criterion (AICC)\citep{1974ITAC...19..716A}.

Having obtained the fits in this section, we can go back to the exercise in Section~\ref{sec:shuffling} and ask whether mock halo properties sampled with equation~\ref{eq:probdist} will recover the assembly bias seen in 2-pt correlation function in Figure~\ref{fig:2ptcorr}. In Figure~\ref{fig:2ptcorrproof}, we show the ratio of 2-pt correlation from such mocks to the 2-pt correlation from the original sample and we can see that there is 10-15\% agreement of mocks with the simulation except for small separations. This reinforces the ability of the mock-making algorithm to reproduce assembly bias at a given redshift between z=0-4.
\section{Improving the dynamic range of a Large Volume Simulation}
\label{sec:demo}
\begin{figure*}
\includegraphics[width=0.8\linewidth,trim=0 0 5 5,clip]{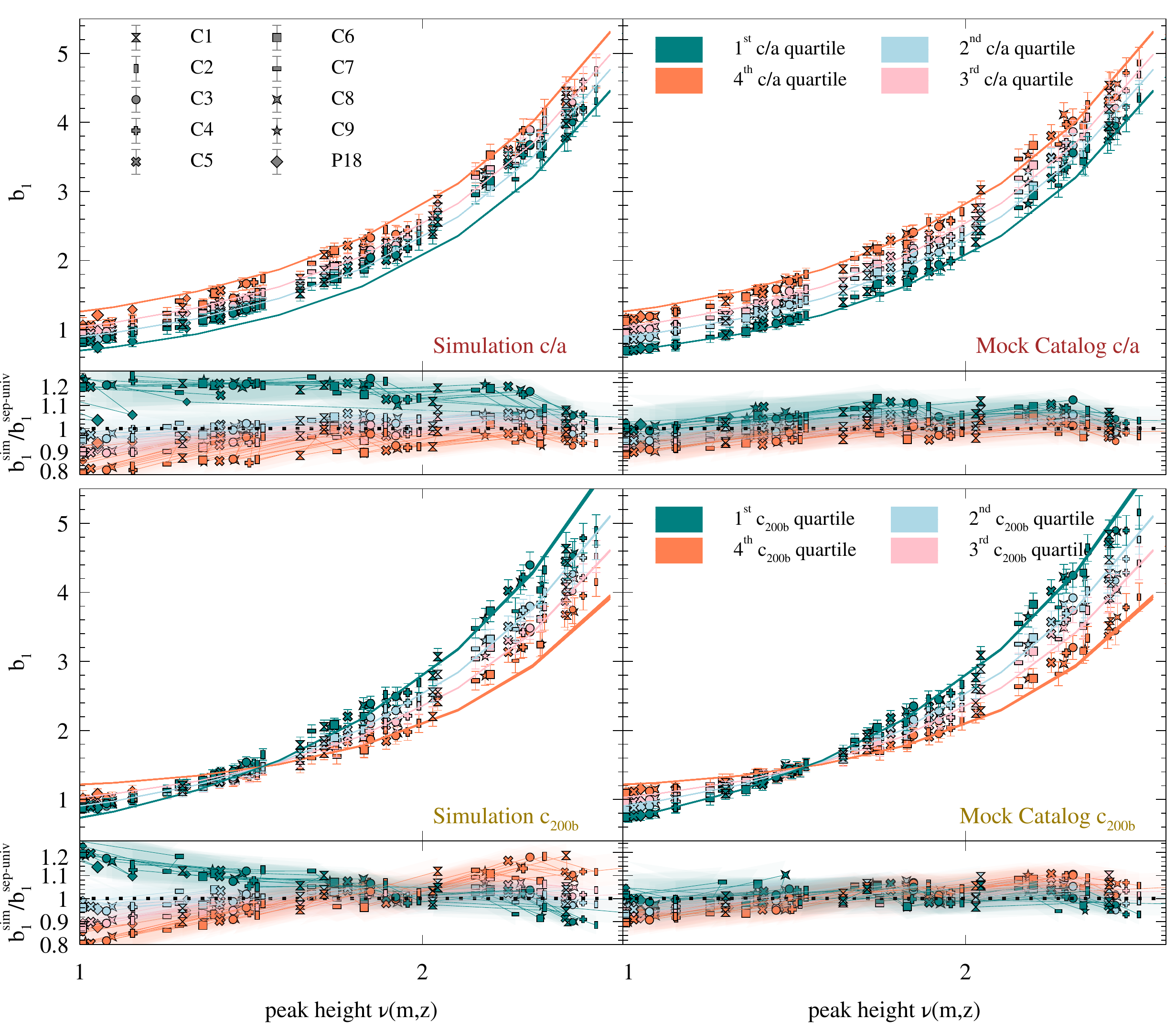} 
\caption{Comparison of assembly bias due to halo shape $c/a$ \emph{(top row)} and concentration $c_{\rm 200b}$ \emph{(bottom row)} in our large-volume, low-resolution simulations and corresponding mocks with Separate Universe (SU) calibrations. For each halo property $c$, the four different coloured data markers show the average halo bias in the four quartiles of $c$, with values $c$ measured directly in the low-resolution simulation \emph{(left panels)} and assigned by our conditional sampling algorithm \emph{(right panels)}. Each type of marker denotes a different cosmology as indicated in the legend (See Table~\ref{tab:sims} and Figure~\ref{fig:cosm-param-space}) and the different redshift data are denoted by different marker sizes (z=0-4). The solid curves, repeated for each property in the corresponding left and right panel, show the calibration for assembly bias using the SU technique. The small lower sub-panels in each case show the ratio of each assembly bias curve with the SU calibration. Low-mass haloes resolved with $\lesssim500$ particles in the simulation (left hand panels) fail to reproduce the full strength of assembly bias, while corresponding haloes in the mock catalogs (right hand panels) perform much better down to a 30 particle threshold.}

\label{fig:largevolmocksu1}
\end{figure*}
In this section, we take the largest simulations available and demonstrate how we can use the mock-making algorithm to provide estimates of halo properties to the poorly resolved haloes at the lower mass end of the simulation. These low-mass haloes ($\leq 500$ particles) have the distribution of their properties offset from those extracted from higher resolution simulation (See Figure B1-B2 of \citet{2021MNRAS.503.2053R}). Our mock estimates not only improve the small-scale properties of these low mass haloes but also large-scale properties correlated with these small scales such as the two-point correlation function and the large-scale bias\footnote{The effect of poor resolution on halo assembly bias is demonstrated in Figure B3 of \citet{2021MNRAS.503.2053R} }. We take the set of simulations which have a box-size of $600 \Mpch$: 9 cosmologies from Suite-III and a single Planck 18 cosmology simulation from Suite -I. We will sample equation~\ref{eq:probdist} for the poorly resolved haloes that have a total particle count of 30-500 particles, improvements upon the original simulation are expected in this mass range or particle count. 

\begin{figure*}
\includegraphics[width=0.8\linewidth,trim=0 0 5 5,clip]{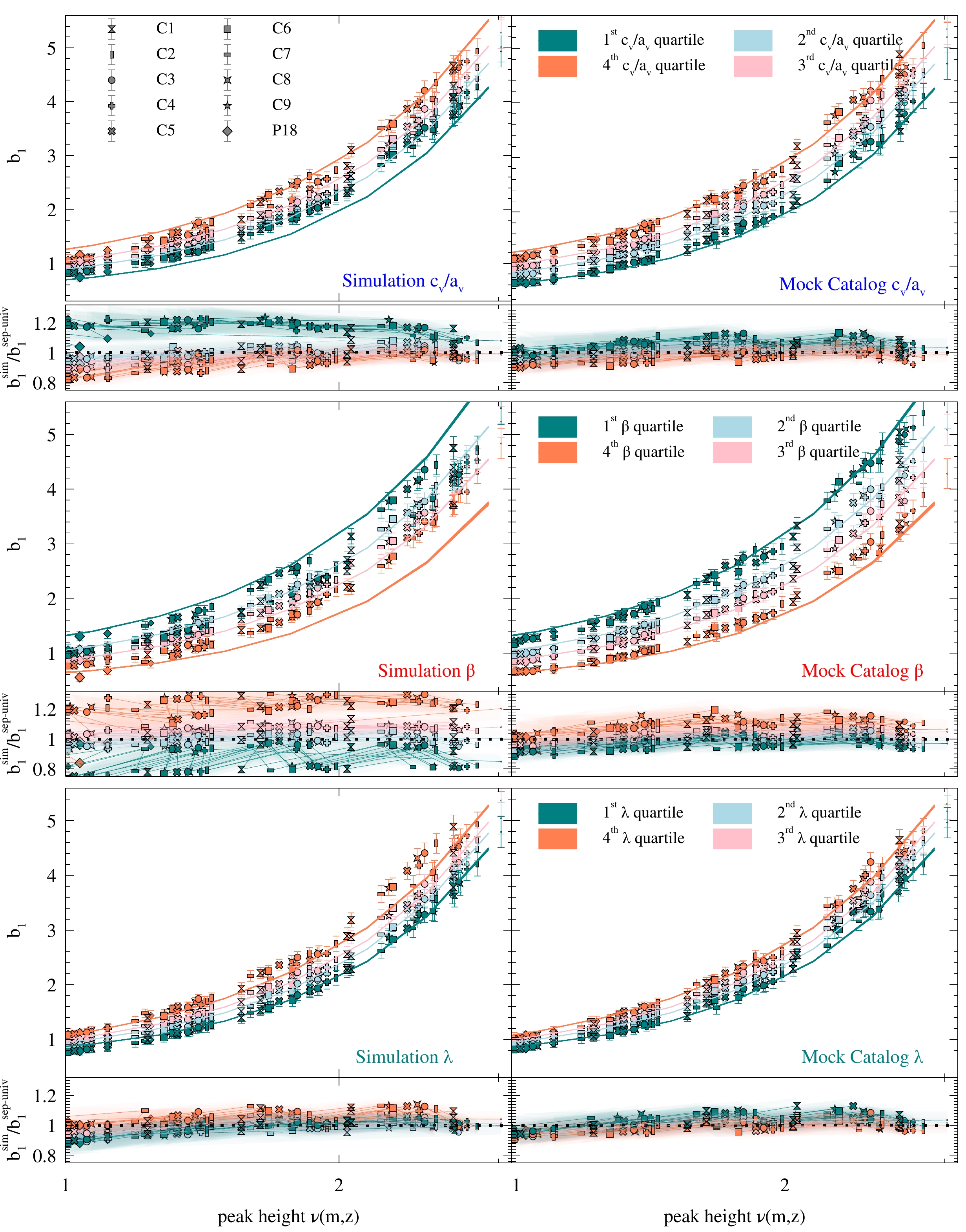} 
\caption{Same as Figure~\ref{fig:largevolmocksu1}, showing results for velocity ellipsoid asphericity $c_{v}/a_{v}$ \emph{(top row)}, velocity anisotropy $\beta$ \emph{(middle row)} and halo spin $\lambda$ \emph{(bottom row)}. Similarly to halo shape and concentration, we again see that our mock algorithm outperforms the actual simulation in reproducing assembly bias trends for poorly resolved haloes.}
\label{fig:largevolmocksu2}
\end{figure*}

The first step is to estimate mass and tidal anisotropy $\tilde{\alpha}$ as described in Section~\ref{sec:methods}. These are not affected by the resolution effects unlike the other halo properties and are well resolved for haloes having more than 30 particles (see last panel of Figure B1-B2 of \citet{2021MNRAS.503.2053R}). Using $m$ and $\tilde{\alpha}$ values of a halo one can sample the equation~\ref{eq:probdist} to obtain its appropriate halo property. It is obvious that sampling a probability distribution is not going to give accurate results on a halo by halo basis. However, our focus is on aiding situations where we require to work with a large number of haloes and are interested in large-scale statistics rather than getting individual properties right. A comparison of the inherent correlations that exist between the large scale bias and small scale halo property in the two sets of halo properties i.e, our mocks and \textsc{rockstar} halo properties, will demonstrate the extent of improvement our algorithm can provide over the conventional halo finding method.  
The markers in  Figure~\ref{fig:largevolmocksu1} and~\ref{fig:largevolmocksu2} show the large-scale bias of the halos belonging to different quartiles of the halo properties generated using our mocks (right panel) and using the conventional halo finder \textsc{rockstar} (left panel) as a function of peak height. The difference in the value of bias in the different quartiles of the same property shows the existence of correlations between the halo properties and large-scale bias or otherwise known as halo assembly bias. The peak height $\nu$ has been useful to superpose the results from different redshifts \citep{2007MNRAS.377L...5G} and on cosmology \citep{2021MNRAS.507.3412C} which is consistent with our findings.
Both the panels show assembly bias with minor differences.
To quantify these minor differences we overlay, as a standard reference, the computation of bias using the Separate Universe (SU) technique (solid lines)\footnote{\label{footnote:SUcaveat}It can be seen from equation 27 of \citet{2020MNRAS.499.4418R} that the SU calibrations require fits from simulation for the distribution of $\alpha$. They have used haloes with $\geq 400$ particles, which are, for this purpose, well resolved. This results in well-tested calibrations spanning the peak height from 1.1 to 2.8}. The SU technique is more accurate as it uses the peak background split approach to compute halo bias and hence unaffected by the limitations that come with working in a finite volume \citep{2020MNRAS.499.4418R}. The smaller panel at the bottom gives the ratio of bias from the simulation compared to the SU bias. The different markers both in the main bias plot and the smaller ratio plot represent different redshifts as given by the label. Since we exclusively show results with these small mass / poorly resolved haloes that span larger and larger peak height for increasing redshift, the assembly bias inferred from the conventional method does not match with the SU curve as well as those inferred from our mock catalog at all peak heights. 
These differences are most apparent in the ratio with SU measurements in the bottom sub-panels; the largest errors are in the upper and lower quartile measurements of the halo bias and relatively smaller errors in the other two quartiles. It must be noted that the SU calibrations are well-tested only for peak heights larger than 1.1 (See footnote \ref{footnote:SUcaveat}).  Though it should not be much of an issue to slightly extend the calibrations to lower peak heights (since the SU calibrations in the right panel of Figure 3 of \citet{2020MNRAS.499.4418R} are slowly varying functions of peak height), we suggest to alternatively consider a comparison between our low-resolution mocks and high-resolution simulation for small peak height. This has been shown in Figure~\ref{fig:comp mock with high-res}

Though each row in Figure~\ref{fig:largevolmocksu1} and ~\ref{fig:largevolmocksu2} shows the comparison of the conventional method with our mocks for five different halo properties (as indicated by the label), the trends discussed so far hold collectively for all the halo properties indicating that this method can possibly be used for other halo properties not included in this analysis. A large number of data points (i.e, 10 cosmologies $\times$ 8 redshifts $\times$ 2 mass bins) leads to a crowded plot resembling a large scatter. However, point-by-point comparison of the left and right panels reveals that mocks tend to agree with the SU calibrations with a maximum of 5\%(10\%) deviations for the inner(outer) quartiles while the halo finder mocks tend to have larger deviations 10\%(20\%) for the inner (outer) quartiles. 

For a single configuration of box size $600\Mpch$, $1024^3$ particles and Planck18 cosmology, we estimated order of magnitude gain in mass and halo number density at $z=0$ (as previously noted by \citet{2021MNRAS.503.2053R} for a similar size box of WMAP cosmology). For larger redshifts, we get higher gains in number density i.e., two orders of magnitude at $z=2$, three orders of magnitude at $z=4$ (see vertical dashed line in Figure~\ref{fig:gain}). To put it another way, we can also say that our algorithm opens accessibility to an entire snapshot of the simulation (around a hundred thousand haloes) given that there was only tens of well-resolved haloes at $z=4$ in our box. Because of the steep fall in the halo mass function at larger masses, one can expect bigger gains in number density for larger Gpc-sized boxes with the same $N_{\rm part}$ (see Appendix~\ref{app:gain}).

\section{Summary}
\label{sec:summary}
Here we have explored, at various redshifts and cosmologies, how the knowledge of a halo's mass and environment (tidal anisotropy) can provide sufficient information to improve the statistical accuracy of several of the halo's internal properties namely halo concentration, spin, shape, velocity ellipsoid ratio, velocity anisotropy. This can be useful in a scenario where we have poorly estimated internal properties and simultaneously well-estimated halo mass and environment, for e.g., in the case of low mass haloes in large-volume and low-resolution simulations and light cone catalogs.

The prescription that we provided for generating such mocks in \citet{2021MNRAS.503.2053R} now includes additional redshift and cosmology dependences and can not only produce the correct overall distribution of the internal properties but also correct assembly bias in clustering statistics like the two-point correlation function and large scale bias. This is because our mocks are based on the finding in \citep{2019MNRAS.489.2977R} that tidal anisotropy statistically explains the halo assembly bias with respect to several halo properties.
Our main results are the following,
\begin{itemize}
\item We confirm the applicability of our technique for 8 different redshifts ranging from $z=0$ to $z=4$ (Section~\ref{sec:shuffling}). This is done by shuffling the halo properties within bins of mass and tidal anisotropy and showing that such a shuffled sample recovers the assembly bias in the 2-point correlation function down to $5 \times R_{200b}$ of the smallest halo (Figure~\ref{fig:2ptcorr}). 
\item We refit the parameters of the probability distribution of a halo property c, i.e, $p(c|\tilde{\alpha},m)$ given by equation~\ref{eq:probdist}, now including data from several redshifts (between 0 and 4) and 12 different cosmologies (Figure~\ref{fig:fits} and Table~\ref{table:ca}-\ref{table:c200b}).
We also perform consistency checks (Figure~\ref{fig:comp mock with high-res} and \ref{fig:2ptcorrproof}) to verify the applicability of the fits across several redshifts.
\item We generate mocks with our method in large volume boxes having  different $\Lambda \rm{CDM}$ cosmologies (10 different cosmologies)   and find improvements in halo assembly bias at all redshifts.
\begin{itemize}
    \item Our mocks can be used to provide statistically realistic estimates of halo properties to the poorly resolved haloes having particle resolution between 30-500, thus increasing the dynamic range of the simulation by an order of magnitude (Figure~\ref{fig:largevolmocksu1} and~\ref{fig:largevolmocksu2}). 
    \item For a box of volume $(600 \Mpch)^3$, the estimated gain in number density from using our mocks increases with increasing redshift with a maximum gain of 3 orders of magnitude at $z=4$. The gain in number density will be higher for larger volumes (Figure~\ref{fig:gain}). 
\end{itemize}
\end{itemize}

Our mocks are useful in studying for e.g., effects of halo properties on galaxy-galaxy lensing at large scales, accessing the BAO using less massive haloes as many be a requirement for a particular observational sample. Our mocks have been calibrated in the redshift ranges which are accessible by the upcoming surveys like the Vera Rubin Observatory Legacy Survey of Space and Time  \citep{2018arXiv180901669T} and the Euclid \citep{2011arXiv1110.3193L} and can be used to increase the accessible dynamic range of Gpc scaled mock catalogs \cite{2019ApJ...875...69D,2021MNRAS.506.2659F,2021MNRAS.508.4017M}.

There are several scopes for improvement of our mocks which we will return to in the future; we have not included substructure in our analysis, and bi-spectrum is another interesting large scale quantity whose correlations with small scale properties have not been inspected.
\section*{Acknowledgments}
SR would like to thank Ravi Sheth and Carlo Giocoli for the initial discussions, Aseem Paranjape, Shadab Alam and Pierluigi Monaco for comments on the draft and R. Krishnan for proofreading. We also thank the University Grants Commission (UGC) India, for funding our research and gratefully  acknowledge the use of high performance computing facilities at IUCAA, Pune\footnote{\url{http://hpc.iucaa.in/}}.
We would also like to gratefully acknowledge Lucia A. Perez and the CAMELS team for providing us their simulation snapshots for some of the analysis.
\section*{Data Availability}
No new data were generated in support of this research. The simulations used in this work are available from the authors upon request. 

\bibliography{reference.bib}

\appendix
\section{Correlation between halo concentration and local environment}
\begin{figure*}
 \includegraphics[width=0.99\linewidth]{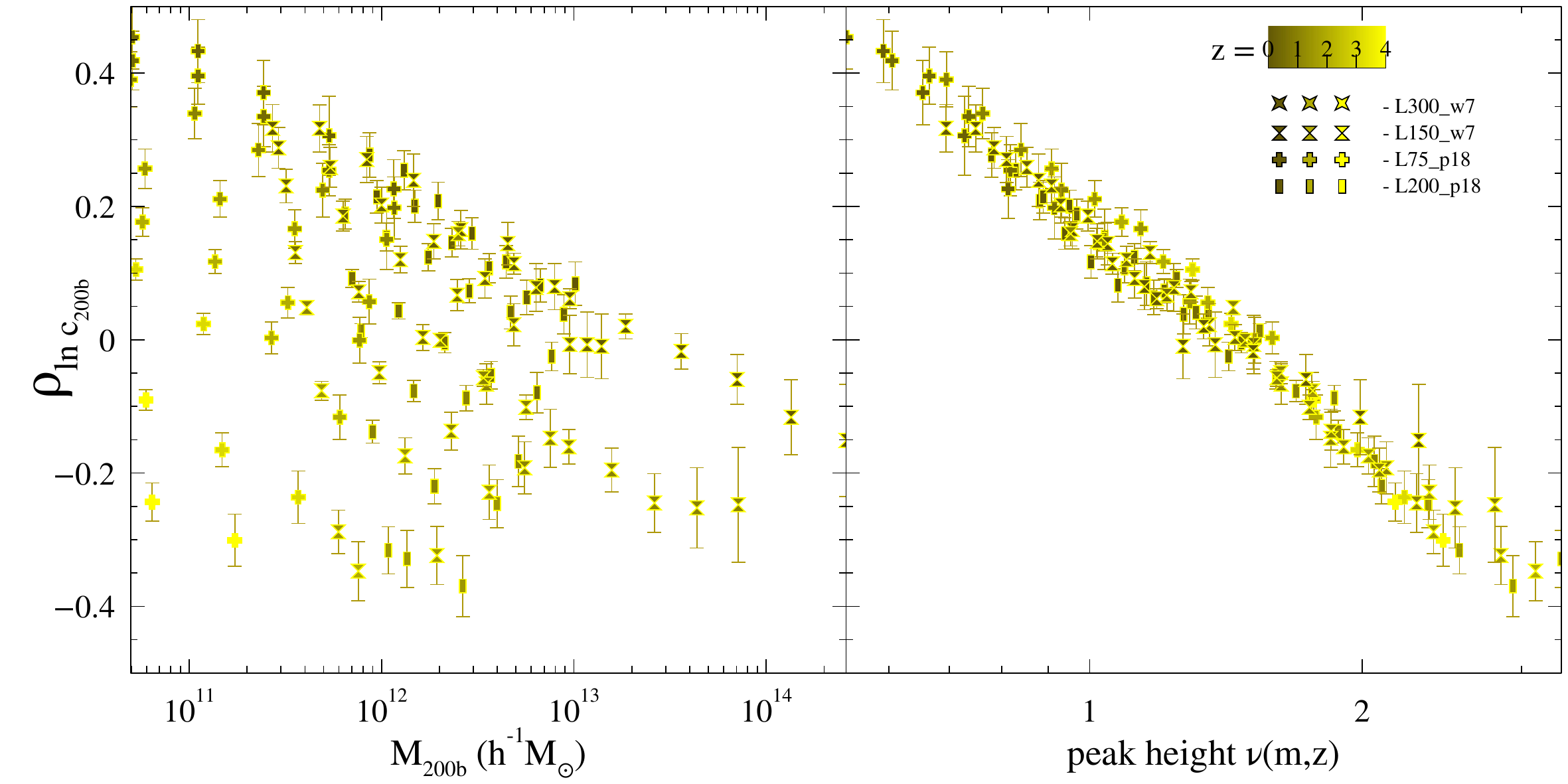}
 \caption{Correlation coefficient of the halo concentration $\ln c_{\rm 200b}$ with standardised tidal anisotropy $\tilde{\alpha}$ as a function of halo mass \emph{(left panel)} and as a function of peak height $\nu$ \emph{(right panel)} for different redshifts ranging from $z=0-4$ as indicated by shades of yellow, different box sizes and two cosmologies (WMPA-7 and Planck-18) as indicated by the markers and labels.  }
 \label{fig:c200brho}
\end{figure*}

It becomes clear from comparing the left and right panels of Figure~\ref{fig:c200brho} that the value of $\rho_{\ln c_{\rm{200b}}}$ for different redshifts/cosmologies is closer to a universal relation when plotted against peak height instead of halo mass; we can fit the relation with a quadratic in $\ln \nu$ and the chisquare analysis gives a pvalue of 0.035 when we allow for an error of $10\%$ of the binned measurement to account for systematic errors like binning, NFW fitting, box size (see Table~\ref{table:c200b}).  

\begin{table}
\centering
 \caption{Best fit coefficients and covariance matrices for the polynomial fits of the mean $\mu_{c}$, variance $\sigma_c$ and Pearson coefficient $\rho_c$ (the correlation with $\tilde{\alpha}$) for the halo property $c=c/a$ obtained from simulations spanning redshifts from 0 to 4 and different cosmologies as shown in Table~\ref{tab:sims}. In each sub-table, the caption shows the polynomial form as a function of $\ln \nu$, the first row gives the best fit values, the second row gives the standard deviation, and the last few rows give the correlation coefficients.}
  \label{table:ca}
 \subcaption{$\mu_{c/a}={(c/a)}_{0}+{(c/a)}_{1}\ln \nu+{(c/a)}_{2}(\ln \nu)^2$}
 \renewcommand{\arraystretch}{1.5} 
 \begin{tabular}{llllll}
 \hline 
 \hline  
&${(c/a)_0}$&${(c/a)_1}$&${(c/a)_2}$&$\chi^2(\rm 132\ d.o.f)$&\\ \hline 
value&0.5537&-0.2080&0.0534&121.50&\\ \hline 
std dev&0.0012&0.0042&0.0059& \\ \hline 
corr ${(c/a)_0}$&1.0000&-0.0741&-0.3124& \\ \hline 
corr ${(c/a)_1}$&-&1.0000&-0.8148& \\ \hline 
 
 \hline \\ 
\end{tabular}
 \subcaption{$\sigma_{c/a}=\sigma^{c/a}_{0}+\sigma^{c/a}_{1}\ln \nu+\sigma^{c/a}_{2}(\ln \nu)^2$}
 \renewcommand{\arraystretch}{1.5} 
 \begin{tabular}{llllll}
 \hline 
 \hline  
&${\sigma^{c/a}_0}$&${\sigma^{c/a}_1}$&${\sigma^{c/a}_2}$&$\chi^2(\rm 132\ d.o.f)$&\\ \hline 
value&0.1276&-0.0060&-0.0152&121.15&\\ \hline 
std dev&0.0004&0.0011&0.0020& \\ \hline 
corr ${\sigma^{c/a}_0}$&1.0000&0.1524&-0.5440& \\ \hline 
corr ${\sigma^{c/a}_1}$&-&1.0000&-0.7026& \\ \hline 
 
 \hline \\ 
\end{tabular}
  \subcaption{$\rho_{c/a}=\rho^{c/a}_{0}+\rho^{c/a}_{1} \ln \nu$}
 \renewcommand{\arraystretch}{1.5} 
 \begin{tabular}{lllll}
 \hline 
 \hline  
&${\rho^{c/a}_0}$&${\rho^{c/a}_1}$&$\chi^2(\rm 133\ d.o.f)$&\\ \hline 
value&0.1978&-0.0171&120.78&\\ \hline 
std dev&0.0013&0.0035& \\ \hline 
corr ${\rho^{c/a}_0}$&1.0000&-0.2412& \\ \hline 
 
 \hline \\ 
\end{tabular}
\end{table}

\begin{table}
\centering
 \caption{Best fit coefficients and covariance matrices for $c_{v}/a_{v}$.}
  \label{table:vca}
 \subcaption{$\avg{c_{v}/a_{v}}={(c_{v}/a_{v})}_{0}+{(c_{v}/a_{v})}_{1}\ln \nu$}
 \renewcommand{\arraystretch}{1.5} 
 \begin{tabular}{lllll}
 \hline 
 \hline  
&${(c_{v}/a_{v})_0}$&${(c_{v}/a_{v})_1}$&$\chi^2(\rm 133\ d.o.f)$&\\ \hline 
value&0.8112&-0.0990&23.26&\\ \hline 
std dev&0.0016&0.0036& \\ \hline 
corr ${(c_{v}/a_{v})_0}$&1.0000&-0.5126& \\ \hline 
 
 \hline \\ 
\end{tabular}
 \subcaption{$\sigma^{c_{v}/a_{v}}=\sigma^{c_{v}/a_{v}}_{0}+\sigma^{c_{v}/a_{v}}_{1}\ln \nu+\sigma^{c_{v}/a_{v}}_{2}(\ln \nu)^2$}
 \renewcommand{\arraystretch}{1.5} 
 \begin{tabular}{llllll}
 \hline 
 \hline  
&${\sigma^{c_{v}/a_{v}}_0}$&${\sigma^{c_{v}/a_{v}}_1}$&${\sigma^{c_{v}/a_{v}}_2}$&$\chi^2(\rm 132\ d.o.f)$&\\ \hline 
value&0.0740&0.0247&-0.0168&23.85&\\ \hline 
std dev&0.0008&0.0024&0.0041& \\ \hline 
corr ${\sigma^{c_{v}/a_{v}}_0}$&1.0000&0.2087&-0.5486& \\ \hline 
corr ${\sigma^{c_{v}/a_{v}}_1}$&-&1.0000&-0.7096& \\ \hline 
 
 \hline \\ 
\end{tabular}
  \subcaption{$\rho^{c_{v}/a_{v}}=\rho^{c_{v}/a_{v}}_{0}+\rho^{c_{v}/a_{v}}_{1}\ln \nu$}
 \renewcommand{\arraystretch}{1.5} 
 \begin{tabular}{lllll}
 \hline 
 \hline  
&${\rho^{c_{v}/a_{v}}_0}$&${\rho^{c_{v}/a_{v}}_1}$&$\chi^2(\rm 133\ d.o.f)$&\\ \hline 
value&0.1985&0.0607&112.04&\\ \hline 
std dev&0.0012&0.0034& \\ \hline 
corr ${\rho^{c_{v}/a_{v}}_0}$&1.0000&-0.1894& \\ \hline 
 
 \hline \\ 
\end{tabular}
\end{table}

\begin{table}
\centering
 \caption{Best fit coefficients and covariance matrices for $\ln (1-\beta)$.}
  \label{table:beta}
 \subcaption{$\mu_{\ln (1-\beta)}={\beta}_{0}+{\beta}_{1}\ln \nu$}
 \renewcommand{\arraystretch}{1.5} 
 \begin{tabular}{lllll}
 \hline 
 \hline  
&${\beta_0}$&${\beta_1}$&$\chi^2(\rm 133\ d.o.f)$&\\ \hline 
value&-0.2105&-0.0642&45.24&\\ \hline 
std dev&0.0020&0.0052& \\ \hline 
corr ${\beta_0}$&1.0000&-0.2881& \\ \hline 
 
 \hline \\ 
\end{tabular}
 \subcaption{$\sigma_{\ln (1-\beta)}=\sigma^{\beta}_{0}+\sigma^{\beta}_{1}\ln \nu$}
 \renewcommand{\arraystretch}{1.5} 
 \begin{tabular}{lllll}
 \hline 
 \hline  
&${\sigma^{\beta}_0}$&${\sigma^{\beta}_1}$&$\chi^2(\rm 133\ d.o.f)$&\\ \hline 
value&0.1744&0.0089&69.70&\\ \hline 
std dev&0.0004&0.0011& \\ \hline 
corr ${\sigma^{\beta}_0}$&1.0000&-0.3345& \\ \hline 
 
 \hline \\ 
\end{tabular}
  \subcaption{$\rho_{\ln (1-\beta)}=\rho^{\beta}_{0}+\rho^{\beta}_{1}\ln \nu+\rho^{\beta}_{2}(\ln \nu)^2$}
 \renewcommand{\arraystretch}{1.5} 
 \begin{tabular}{llllll}
 \hline 
 \hline  
&${\rho^{\beta}_0}$&${\rho^{\beta}_1}$&${\rho^{\beta}_2}$&$\chi^2(\rm 132\ d.o.f)$&\\ \hline 
value&0.2368&0.1011&0.1282&112.10&\\ \hline 
std dev&0.0018&0.0048&0.0091& \\ \hline 
corr ${\rho^{\beta}_0}$&1.0000&0.1929&-0.5790& \\ \hline 
corr ${\rho^{\beta}_1}$&-&1.0000&-0.5978& \\ \hline 
 
 \hline \\ 
\end{tabular}
\end{table}

\begin{table}
\centering
 \caption{Best fit coefficients and covariance matrices for $\ln \lambda$.}
  \label{table:spin}
 \subcaption{$\mu_{\ln \lambda}={\lambda}_{0}+{\lambda}_{1}\ln \nu+{\lambda}_{2}(\ln \nu)^2$}
 \renewcommand{\arraystretch}{1.5} 
 \begin{tabular}{llllll}
 \hline 
 \hline  
&${{\lambda}_0}$&${{\lambda}_1}$&${{\lambda}_2}$&$\chi^2(\rm 228\ d.o.f)$&\\ \hline 
value&-3.3372&-0.1728&-0.1305&18.48&\\ \hline 
std dev&0.0059&0.0153&0.0257& \\ \hline 
corr ${{\lambda}_0}$&1.0000&0.1902&-0.5694& \\ \hline 
corr ${{\lambda}_1}$&-&1.0000&-0.6927& \\ \hline 
 
 \hline \\ 
\end{tabular}
 \subcaption{$\sigma_{\ln \lambda}=\sigma^{\lambda}_{0}$}
 \renewcommand{\arraystretch}{1.5} 
 \begin{tabular}{llll}
 \hline 
 \hline  
&${\sigma^{\lambda}_0}$&$\chi^2(\rm 134\ d.o.f)$&\\ \hline 
value&0.5808&58.20&\\ \hline 
std dev&0.0013& \\ \hline 
 
 \hline \\ 
\end{tabular}
  \subcaption{$\rho_{\ln \lambda}=\rho^{\lambda}_{0}+\rho^{\lambda}_{1} \ln \nu+\rho^{\lambda}_{2} (\ln \nu)^2$}
 \renewcommand{\arraystretch}{1.5} 
 \begin{tabular}{llllll}
 \hline 
 \hline  
&${\rho^{\lambda}_0}$&${\rho^{\lambda}_1}$&${\rho^{\lambda}_2}$&$\chi^2(\rm 228\ d.o.f)$&\\ \hline 
value&0.0711&0.1414&-0.0494&210.16&\\ \hline 
std dev&0.0014&0.0030&0.0065& \\ \hline 
corr ${\rho^{\lambda}_0}$&1.0000&0.3433&-0.6752& \\ \hline 
corr ${\rho^{\lambda}_1}$&-&1.0000&-0.4210& \\ \hline 
 
 \hline \\ 
\end{tabular}
\end{table}

\begin{table}
\centering
 \caption{Best fit coefficients and covariance matrix for $\ln c_{200b}$ \\ $\rho_{\ln c_{200b}}=\rho^{c_{200b}}_{0}+\rho^{c_{200b}}_{1} \ln \nu+\rho^{c_{200b}}_{2} (\ln \nu)^2$.}
  \label{table:c200b}
 \renewcommand{\arraystretch}{1.5} 
 \begin{tabular}{llllll}
 \hline 
 \hline  
&${\rho^{c_{200b}}_0}$&${\rho^{c_{200b}}_1}$&${\rho^{c_{200b}}_2}$&$\chi^2(\rm 132\ d.o.f)$&\\ \hline 
value&0.1712&-0.4188&-0.1121&179.34&\\ \hline 
std dev&0.0029&0.0121&0.0197& \\ \hline 
corr ${\rho^{c_{200b}}_0}$&1.0000&-0.2100&-0.2145& \\ \hline 
corr ${\rho^{c_{200b}}_1}$&-&1.0000&-0.7809& \\ \hline 
 
 \hline \\ 
\end{tabular}
\end{table}
\section{Gain in number density of haloes for larger volume mocks}
\label{app:gain}
\begin{figure}
    \centering
    \includegraphics[width=0.99\linewidth]{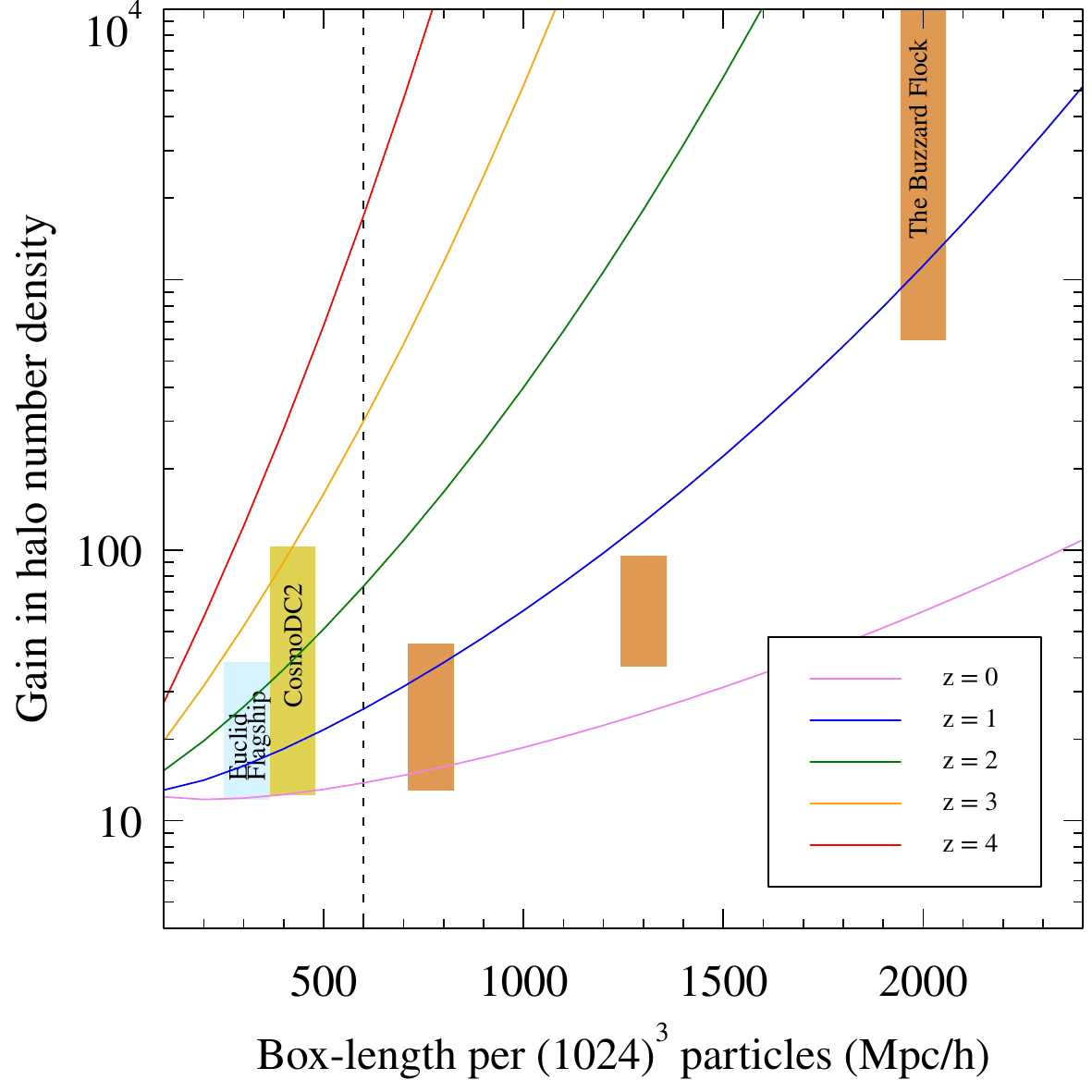}
    \caption{Gain in halo number density for different sized simulations at different redshifts ranging from $z=0-4$. This has been calculated using the \citet{2008ApJ...688..709T} halo mass function with Planck-18 cosmological parameters. The vertical dashed line marks the gain for our current large volume simulation ($600 \Mpch$). We can see that our algorithm has the largest gain for mocks that are either at high redshifts or large volumes at fixed $N_{\rm part}$. The vertical strips with labels show the gains which can be achieved by in different galaxy mocks that are available for different surveys spanning various redshifts (see text for a discussion).}
    \label{fig:gain}
\end{figure}

In this section, the halo mass function by \citet{2008ApJ...688..709T} is used to compute the gain in number density by using our algorithm at different redshifts as well as different box sizes. We will define the gain as the ratio of the number density of haloes after applying our method (i.e., including the $30-500$ particle haloes) to the number density of haloes without the method (i.e., excluding the $30-500$ particle haloes).  Figure~\ref{fig:gain} shows the gain in number density for different large-volume simulations with different box sizes and Planck-18 cosmology. We can visually infer from this figure that the mocks have larger gains and hence are particularly useful for higher redshifts and larger box sizes at fixed $N_{\rm part}$. The vertical dashed line in Figure~\ref{fig:gain} shows the gain for our simulation box used in this work ($600 \Mpch$ from Suite-I). For a comparison with the mocks available in the literature, we have also shown rectangular labels that mark the gains that are achievable at different redshifts in the Buzzard catalogs developed for DES \citep{2019arXiv190102401D}, the flagship galaxy mock catalog of the Euclid Consortium \citep{2017ComAC...4....2P,2022A&A...662A..93E} and CosmoDC2 that supports LSST \citep{2019ApJS..245...26K}.
\section{Consistency checks}
In this section, we want to perform two tests to check for consistency of our mock making algorithm.

\begin{figure}
    \centering
    \includegraphics[width=0.99\linewidth]{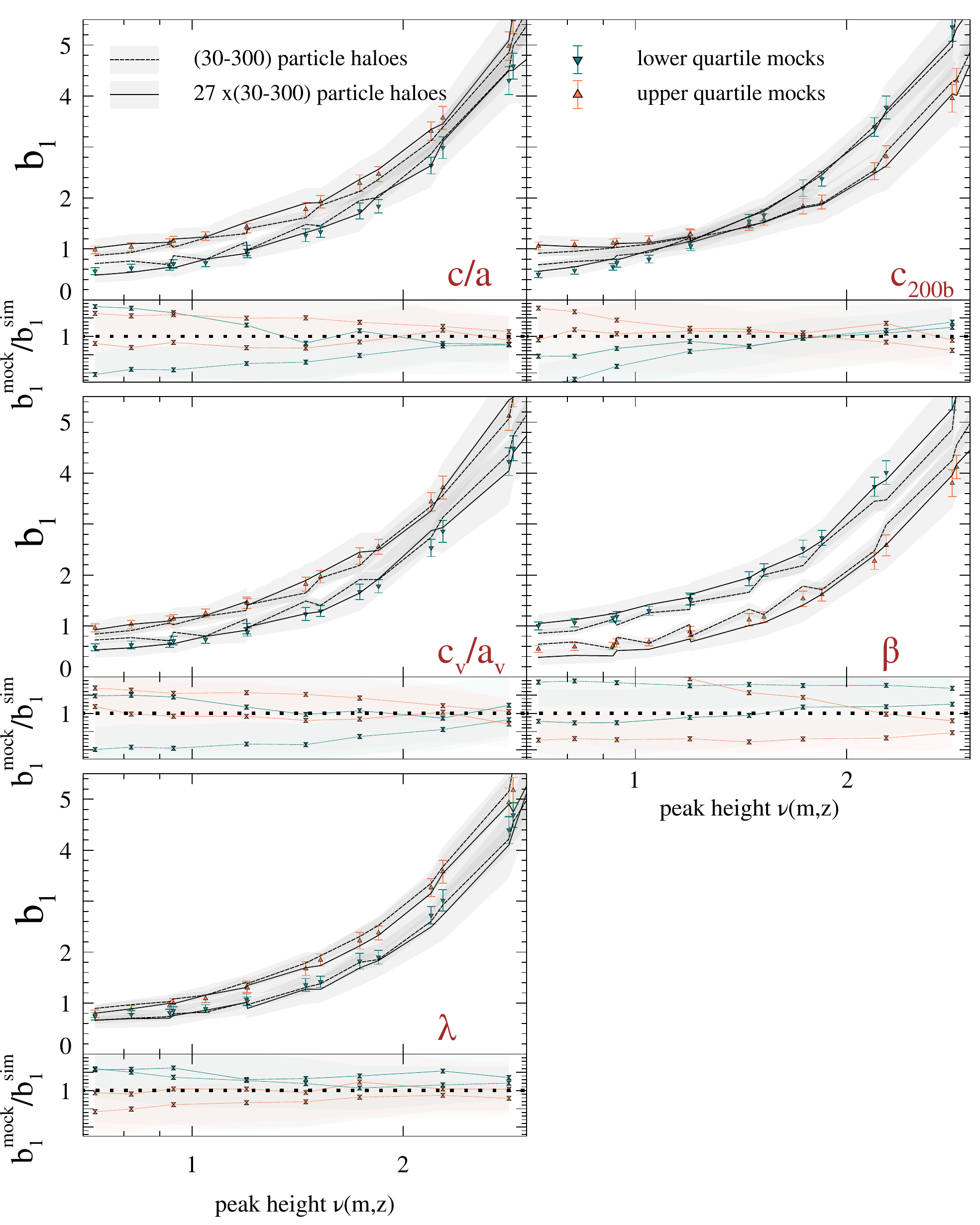}
    \caption{\textbf{Comparison of low resolution mocks (30-300 particle haloes) with high/low-resolution simulation.} Each panel corresponds to a different halo property as given by the label. Here, the orange and green markers denote the bias of haloes in the upper and lower quartile of a halo property sampled using equation 4. The solid(dashed) lines denote bias as obtained from high(low)-resolution simulations after using a halo finder like \textsc{rockstar}. The bottom subplot shows the ratio of bias from mock samples to the bias obtained from high-resolution(solid lines) or low-resolution(dashed lines). We can see that in most cases, the mock sample agrees with the high-resolution simulation better than with the low-resolution simulation.}
    \label{fig:comp mock with high-res}
\end{figure}

In Figure~\ref{fig:comp mock with high-res}, we use a test set of low-resolution and high-resolution(27 $\times$ particle count of low-res) simulations of the Planck18 cosmology and compare its bias with those obtained from our mocks, we can see that for each halo property, the mock mostly matches the high resolution better than the
low-resolution simulation. Though this is a more direct way of demonstrating the performance of the mock algorithm, it is disadvantaged
because of the large errors in bias b1 measurements, thus we can afford to show only the upper and lower quartiles as opposed to all four quartiles in the analysis with the Separate Universe method. 

\begin{figure*}
\includegraphics[width=0.99\linewidth]{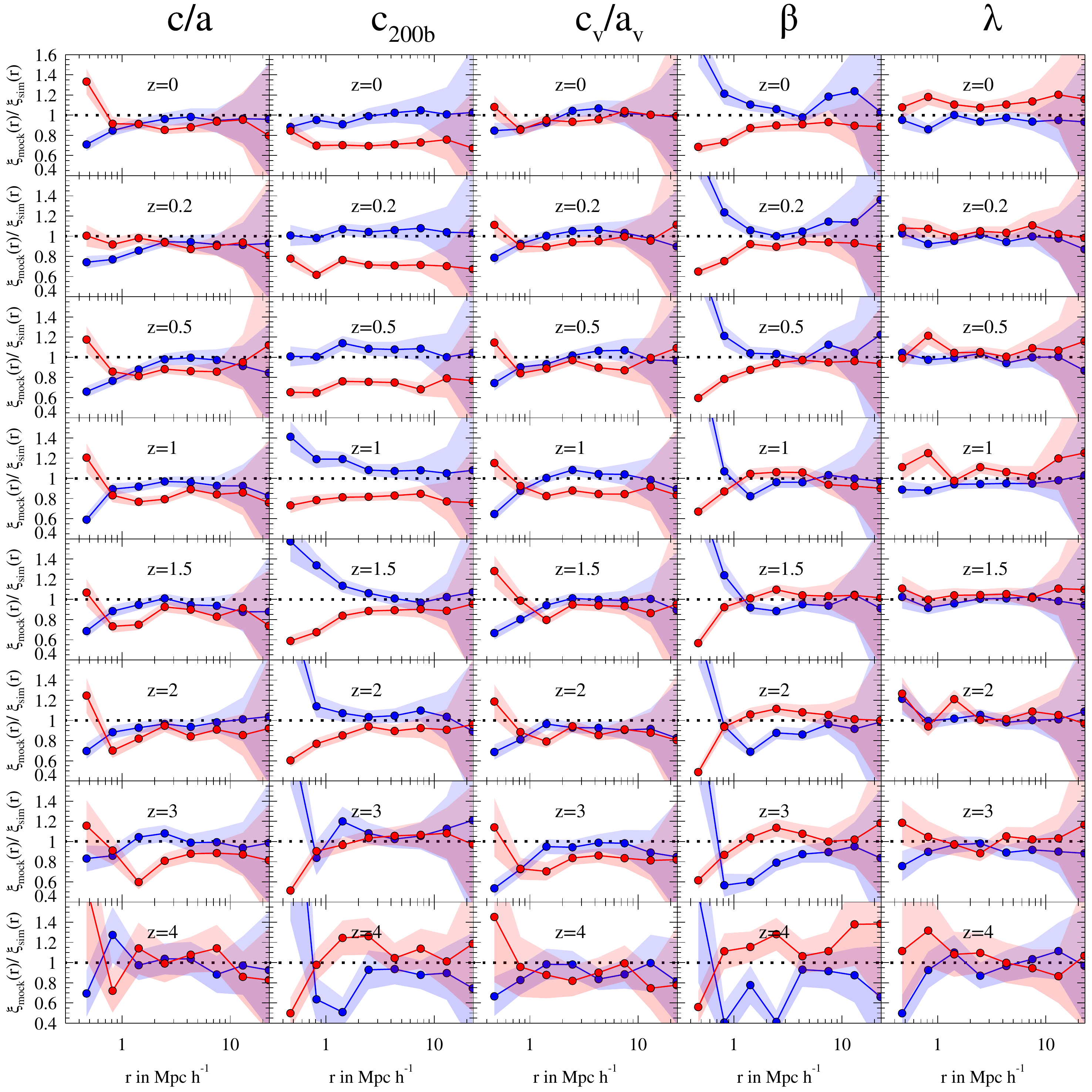}
\caption{Ratio of $\xi(r)$ of the haloes whose properties are sampled using equation 4 (markers)  to the $\xi(r)$ of the same haloes whose properties are measured in the simulations (lines). Each column shows the halo clustering ratio in upper (blue) and lower (red) quartiles of a single halo property and different panels in each column correspond to different redshifts as indicated by the labels. For each quartile of each property, the mock generated using equation~\ref{eq:probdist} reproduces  $\xi(r)$, this can be seen from the ratio of $\xi(r)$ in the shuffled samples to the original sample being mostly consistent with $1$. At separation larger than $r\gtrsim 5R_{\rm 200b}^{min}$ (See Section~\ref{sec:shuffling} for the significance of this scale), the $\xi(r)$ of the mocks and simulations are mostly within $15\%$ of each other though it deviates significantly (upto $40\%$) for a handful of redshifts, separations and halo properties. We have used 125 jackknife samples to obtain errors for samples from $600\Mpch$ box. Here, we use haloes in the range $4\times 10^{11} - 4 \times 10^{12} \Mh$ in the Planck-18 simulation box.}
\label{lastpage}
\label{fig:2ptcorrproof}
\end{figure*}

In another consistency check, we want to see whether sampling halo properties using equation~\ref{eq:probdist} can reproduce the assembly bias in 2-pt correlation function at any redshift. The ratio of the 2-pt correlation obtained from such a sampling (mocks) with the 2-pt correlation from the actual simulation is shown in Figure~\ref{fig:2ptcorrproof}. At very small scales ($r\lesssim 5R_{\rm 200b}^{min}$ ), we don't expect the mocks to reproduce the simulations (see discussion in Section~\ref{sec:shuffling}). At larger separations, we can see that the mocks recover the 2-pt correlation within 15\% for most of the separations with a few exceptions. At redshift $z=4$, where the statistics are worsened by the availability of fewer haloes, the upper quartile of mock $\beta$ has a large deviation (up to $40\%$) from the simulation sample. Another exception is the lower quartile of $c_{200b}$ which deviates by up to ($25\%$) from simulations. Since the 2-pt correlation function of dark matter haloes scales as $b_{1}^2$ at large scales, it can also be expected that the ratios in Figure~\ref{fig:2ptcorrproof} have a larger deviation from $1$ than the ratios in the right small panels of Figure~\ref{fig:largevolmocksu1} and~\ref{fig:largevolmocksu2}.

\end{document}